\documentclass[aps,prl,floatfix,showpacs,twocolumn]{revtex4}
\usepackage{amsmath,amssymb}
\usepackage{graphicx}
\usepackage{epsfig}

\newcommand{\bm}{\boldsymbol}

\begin{document}

\hsize\textwidth\columnwidth\hsize\csname@twocolumnfalse\endcsname

\title{Drude-Interband Coupling, Screening, and the \\ Optical Conductivity of Doped Bilayer Graphene}

\author{Wang-Kong Tse}
\author{A.~H. MacDonald}
\affiliation{Department of Physics, University of Texas, Austin, Texas 78712, USA}

\begin{abstract}
We present a theory of the influence of band renormalization and excitonic electron-electron interaction effects 
on the optical conductivity $\sigma(\omega)$ of doped bilayer graphene. Using the Keldysh formalism, we derive a kinetic equation
from which we extract numerical and approximate analytic results for $\sigma(\omega)$.  Our calculations reveal 
a previously unrecognized mechanism which couples the Drude and interband response and renormalizes the plasmon frequency,
and suggest that screening must play an essential role in explaining the weakly renormalized conductivity seen in recent experiments. 
\end{abstract}
%
\maketitle
\noindent
{\em Introduction}---Experimental progress \cite{Geim} over the past five years has made it possible to isolate  
single-layer graphene (SLG), an atomically two-dimensional electron system based on  
a honeycomb lattice of carbon atoms, and study its electronic properties.  More recently similar techniques have been used to study
multi-layer graphene films and other types of graphitic nanostructures.
One of the surprises in this field is that the electronic properties of Bernal stacked bilayer graphene (BLG) are 
quite distinct \cite{McCann} from those of SLG.  The low-energy electronic excitations of a bilayer are 
\textit{massive} and have momentum-space Berry phase $2\pi$, while those of SLG are massless 
and have Berry phase $\pi$.  

The optical conductivity $\sigma(\omega)$ of BLG has received a lot of attention, both experimentally and theoretically.
Theoretical studies \cite{NilssonBLG,FalkoBLG,CarbotteBLG,NetoBLG,ZhangBLG} of $\sigma(\omega)$ in BLG have so far entirely 
neglected the interactions effects that are known to be crucial in the optical response of regular semiconductors \cite{Koch,Zimmermann}.
The non-interacting electron interpretation of $\sigma(\omega)$ data \cite{WangBLG,BasovBLG,KuzmenkoBLG,CrommieBLG} does
nevertheless appear to be generally successful, surprisingly so since BLG is generally expected to display stronger 
interaction effects than SLG because of its parabolic band dispersion.
For example band structure renormalizations are expected \cite{PoliniSSC} to be relatively modest in 
SLG, but substantially stronger in the BLG case.  Indeed the interacting electron problem in bilayer graphene poses 
a number of interesting new questions because of its unique massive chiral quasiparticles.  

These circumstances call for 
the theoretical analysis of the influence of interactions on $\sigma(\omega)$ presented in this Letter.
We use a 
quantum kinetic equation (QKE) derived using the Keldysh formalism 
to take  e-e band renormalization and excitonic effects 
into account on an equal footing (thus correctly guaranteeing gauge invariance).
We show that although the energy dispersion of BLG is parabolic, its optical properties  
are very different from those of regular semiconductors or semiconductor bilayers: {\em i)} A new coupling between the Drude ({\em i.e.}, intraband) and 
interband optical transition channels follows from the chirality of the BLG 
band eigenstates; {\em ii)} the Drude-interband coupling (DIC) is responsible for 
a renormalization of the leading order long-wavelength plasmon dispersion; 
and {\em iii)} because of the the chirality structure, screening is responsible for an especially strong suppression of interaction effects.

\noindent{\em Quantum Kinetic Equation}---States near the Fermi level of bilayer graphene are described 
by the two-band envelope function Hamiltonian \cite{McCann} $H = -\epsilon_k\bm{\sigma}\cdot\bm{{n}}$, where $\epsilon_k = k^2/2m$, $\bm{{n}} = (\mathrm{cos}2\phi_k,\mathrm{sin}2\phi_k)$, and $\bm{\sigma}$ is the Pauli matrix vector which acts on layer pseudospin degrees of freedom. [We set $\hbar = 1$ throughout restoring it only in the final expressions for $\sigma(\omega)$]. This Hamiltonian is valid when $v_F k \ll \gamma_1$ where $v_F \simeq 10^6\,\mathrm{ms}^{-1}$ is the quasiparticle velocity of SLG and $\gamma_1 \simeq 0.4 \mathrm{eV}$ \cite{BasovBLG,KusmenkoNewPreprint} is the interlayer hopping amplitude.  (We neglect trigonal warping, which is important only at low densities and energies.) 
Although conduction and valence band eigenenergies have the same quadratic dispersion in regular semiconductors and BLG, 
the eigenfunction properties differ.  In the BLG case the conduction and valence band eigenstates are both  
linear combinations of $\pi$-orbitals, whereas in the regular semiconductor case the two bands have orbitals with 
different atomic character.  We will see that this property alone profoundly alters the $\sigma(\omega)$ theory.  Furthermore bilayer graphene is gapless \cite{gapless}.

To incorporate band renormalization and excitonic effects on an equal footing, we derive a quantum kinetic equation for bilayer 
graphene using the Keldysh formalism and a first order exchange-interaction approximation for the interaction self-energy.
Importantly the interaction term in the envelope function is diagonal \cite{interactionapology} in pseudospin labels at each interaction vertex.
To obtain a kinetic equation, it is customary to employ a Wigner representation in which the relative coordinates $\bm{r} \equiv \bm{r}_1-\bm{r}_2$ and $\tau \equiv t_1-t_2$ in the 
Keldysh Green function \cite{Rammer} are Fourier-transformed to obtain momentum and energy variables $\bm{k}$ and $\varepsilon$, and then perform a 
gradient expansions with respect to the `center-of-mass' coordinates $\bm{R} = (\bm{r}_1+\bm{r}_2)/2$ and $t = (t_1+t_2)/2$. 
The $2 \times 2$ distribution function $f_k$ is obtained \cite{Rammer} by integrating the Keldysh Green function over energy. 
For the case of the bilayer graphene Hamiltonian we find that 
%
\begin{equation}
\frac{\partial f_{\bm {k}}}{\partial t}     
+e\bm{\mathcal{E}}\cdot\frac{\partial f_{\bm{k}}}{\partial \bm{k}}+i\left[-\epsilon_k\bm{{n}}\cdot\bm{\sigma}+\Sigma_{\bm{k}},f_{\bm{k}}\right] = 0,
\label{eq2}
\end{equation}
where 
\begin{equation}
\label{eq:Sigma}
\Sigma_{\bm{k}} = -\sum_{\bm{k'}} V_{\bm{k}-\bm{k'} }f_{\bm{k'}} 
\end{equation} 
is the quasiparticle exchange self-energy.  The property that the $2 \times 2$ self-energy matrix at one wavevector is simply an interaction weighted average of distribution function matrices at different wavevectors is a consequence of the model's pseudospin independent interactions.

We consider linear response to an {\em ac} electric field, $\bm{\mathcal{E}} = \bm{\mathcal{E}}_0 e^{-i\omega t}$ and 
write $f_k = f_k^{(0)}+f_k^{(1)}$ where $f_k^{(0)} = (1/2)\sum_{\mu = \pm}n_F(1-\mu\bm{\sigma}\cdot\bm{n})$ is the equilibrium distribution function.
Here $n_F$ is the Fermi function at zero temperature.  
Using Eq.(~\ref{eq2}) we find that 
\begin{eqnarray}
&&-i\omega f_k^{(1)}-i\left[\epsilon_k\bm{\sigma}\cdot\bm{n}+\Sigma_k,f_k^{(1)}\right] = \nonumber \\
&&S_k+i\sum_{k'}V_{k-k'}\left[f_{k'}^{(1)},f_k^{(0)}\right]. 
\label{eq3}
\end{eqnarray}
where $\Sigma_k$ now refers to the self-energy evaluated using $f_k = f_k^{(0)}$ in Eq.~(\ref{eq:Sigma}), and 
\begin{widetext} 
\begin{equation}
S_k = -(e\bm{\mathcal{E}}\cdot\bm{\hat{k}}/2)\sum_{\mu = \pm}\left[\partial n_F(\xi_{k\mu})/\partial k\right]\left(1-\mu\bm{\sigma}\cdot\bm{n}\right) 
+(1/k)\sum_{\mu = \pm}\mu n_F(\xi_{k\mu})\left(e\bm{\mathcal{E}}\times\bm{\hat{k}}\right)\cdot\left(\bm{\sigma}\times\bm{n}\right), \label{eq4}
\end{equation}
is the driving term of the QKE.  The first term in Eq.~(\ref{eq4}) drives intraband transitions and the second term interband transitions. 
The second term on the right-hand side of Eq.~(\ref{eq3}) accounts for changes in the self-energy in the non-equilibrium state. 
Because of this term, Eq.~(\ref{eq3}) is an integral equation which can only be solved numerically. 
The distribution function can also be expressed as a sum of intraband and interband contributions:
\begin{equation}
f_k^{(1)} = \label{eq5}  
(\bm{\mathcal{E}}\cdot\bm{\hat{k}})\left[A(k)+\bm{\sigma}\cdot\bm{n}B(k)+i\left(\bm{\sigma}\times\bm{n}\right)_zG(k)+\sigma_zH(k)\right] 
+(\bm{\mathcal{E}}\times\bm{\hat{k}})_z\left[i\left(\bm{\sigma}\times\bm{n}\right)_zC(k)+\sigma_zD(k)+\bm{\sigma}\cdot\bm{n}E(k)+F(k)\right],
\end{equation} 
where the $\bm{1}, \bm{\sigma}\cdot\bm{n}, (\bm{\sigma}\times\bm{n})_z$, and $\sigma_z$ components of each contribution 
capture respectively changes in total density, conduction {\em vs.} valence band density difference, interlayer coherence, and layer polarization. 
%
%
\end{widetext} 

Substituting Eq.~(\ref{eq5}) into Eq.~(\ref{eq3}) yields a set of eight coupled equations.  
We find that $E,F,G$ and $H$ are all identically zero, that $A(k) = -B(k) = (ie/2\omega)\delta(k-k_F)$, and that 
$C(k)$ and $D(k)$ satisfy the following set of coupled integral equations:
\begin{eqnarray}
\omega C(k)+\delta_k D(k) &=& \theta(k-k_F)\left[-\frac{e}{k}+ \delta \Sigma^{z}(k)\right],\label{eq7} \\
\delta_k C(k)+\omega D(k) 
&=& \theta(k-k_F)\left[\delta \Sigma^{\phi,1}(k) + \delta \Sigma^{\phi,2}(k) \right]. 
\label{eq8}
\end{eqnarray}
where $\delta_k = 2\epsilon_k+\Sigma_{k+}-\Sigma_{k-}$ is the energy needed to create a vertical interband excitation, 
\begin{equation}
\Sigma_{k\mu} = -\sum_{\lambda = \pm}\sum_{k'}V_{k-k'}\theta(k_F-\lambda k')(1+\mu\lambda\mathrm{cos}\,2\phi_{k'k})/2,
\label{eq9}
\end{equation}
is the equilibrium self-energy in band $\mu$, $\phi_{k'k} = \phi_{k'}-\phi_{k}$, and the non-equilibrium self-energy changes are
%
\begin{eqnarray}
\delta \Sigma^{z}(k) &=& \sum_{k'}V_{k-k'}\mathrm{cos}\,\phi_{k'k}D(k'), \label{eq11} \\
\delta \Sigma^{\phi,1}(k) &=& \sum_{k'}V_{k-k'}\mathrm{cos}\,\phi_{k'k}\,\mathrm{cos}\,2\phi_{k'k}C(k'), \label{eq10} \\
\delta \Sigma^{\phi,2}(k) &=& -i\sum_{k'}V_{k-k'}\mathrm{sin}\,\phi_{k'k}\,\mathrm{sin}\,2\phi_{k'k}B(k'). \label{eq12} 
\end{eqnarray}
%
%
Eqs.~(\ref{eq7})-(\ref{eq8}) are the equations of motion for the interlayer coherence $C(k)$ and layer polarization $D(k)$ 
components of the distribution function and describe precession of valence-band pseudospins in 
effective magnetic fields due to the band-energy separation ($\delta_k$) and to 
non-equilibrium self-energy corrections which favor interband coherence ($\delta\Sigma^{\phi,1}$) and layer polarization 
($\delta\Sigma^{z}$).  The second contribution to $\delta\Sigma^{\phi}$ Eq.~(\ref{eq12}) couples Drude and interband response (DIC).
This DIC mechanism is one of the principle results of this Letter.  It appears because the Drude conduction band Fermi surface oscillation 
in an {\em ac} electric field changes the exchange potential experienced by precessing valence-band pseudospins outside the Fermi surface.

The current can be evaluated from the perturbed distribution function using 
$\bm{J} = g_vg_s \, e \, \mathrm{Tr}[\sum_k(1/2)\{\bm{j}_k,f_k^{(1)}\}]$, where $g_v g_s = 4$ is the product of the valley and spin degeneracies  and 
$\bm{j}_k = \partial H/\partial \bm{k} = -(k/m)[(\bm{\sigma}\cdot\bm{\hat{k}})\,\bm{\hat{x}}-(\bm{\sigma}\times\bm{\hat{k}})_z\,\bm{\hat{y}}]$ is the current operator. It then follows from Eq.~(\ref{eq5}) that the conductivity  
\begin{equation}\sigma(\omega) = -(2e/\pi m)\int_0^{\infty}\mathrm{d}k k^2\left[B(k)+iC(k)\right]. 
\label{eq13}
\end{equation}

\noindent{\em Renormalization of Drude Weight and Plasmon Frequency}----Before discussing our general results we comment on the influence of 
DIC on the Drude weight.  At low frequencies the non-interacting response is the   
out-of-phase oscillations of the Fermi surface with respect to the electric field which is captured by $B(k) \propto i/\omega$.
Because of the self-energy correction $\delta \Sigma^{\phi,2}$, there is also an interband [$C(k)$] 
response with the same frequency dependence [See Eq.~(\ref{eq8})]. When a momentum relaxation time $\tau$ is added to the 
theory the $i/\omega$ contribution to the conductivity evolves into a Drude peak contribution proportional to 
$\tau/(1-i\omega \tau)$.  The coefficient of this contribution is known as the Drude weight $\mathcal{D}$.
When DIC is included we find that for bilayer graphene   
%
\begin{equation}
\sigma^{\mathcal{D}}(\omega) = \frac{(2e^2\varepsilon_F/\pi \hbar)\mathcal{\tilde{D}}\; \tau}{1-i\omega \tau}
\label{eq16}
\end{equation}
where the interaction induced Drude weight renormalization is given to leading order in $e^2$ by  
\begin{equation}
\mathcal{\tilde{D}} = 1+\frac{e^2}{2\pi m\varepsilon_F}\int_{k_F}^{\infty}\mathrm{d}k\;\frac{k^2}{\delta_k}\mathcal{R}\left(\frac{k}{k_F}\right). \label{eq17} 
\end{equation}
Here $\mathcal{R}(x) = (4/15x^3)\{(x+1)(x^4-x^2+1)\mathbb{E}[4x/(x+1)^2]-(x^2+1)(x-1)^2(x+1)\mathbb{K}[4x/(x+1)^2]\}$ and 
$\mathbb{K}, \mathbb{E}$ are respectively complete elliptic integrals of the first and second kind. 
 
\begin{figure}[!tb]
  \includegraphics[width=8.5cm,angle=0]{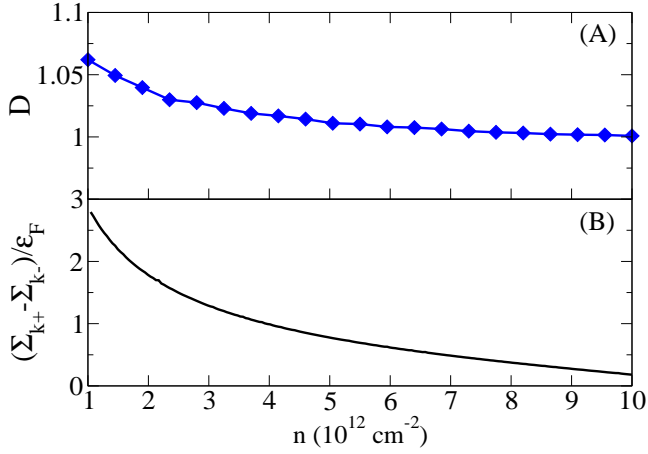} 
\caption{(A) Renormalized Drude weight $\mathcal{\tilde{D}}$ versus density $n$. The enhancement of the Drude weight increases with decreasing density. (B) Band renormalization $(\Sigma_{k+}-\Sigma_{k-})/\varepsilon_F$ at $k = k_F$ versus density.  These results were evaluated with Coulombic electron-electron interactions and dielectric constant 
$\kappa=1$, corresponding to a suspended graphene sample.} 
\label{fig3}
\end{figure}

One important consequence of DIC is renormalization of the plasmon frequency. 
In regular semiconductors with parabolic dispersion, the plasmon frequency $\omega_{\mathrm{p}}$ has no long wavelength interaction renormalization \cite{Kohn} because of 
Galilean invariance.  Since graphene systems are not Galilean invariant, their plasmon frequencies 
are \cite{Vignale} renormalized. Using the well-known relation between the optical conductivity and the polarizability,
$\sigma(\omega) = \mathrm{lim}_{q\to 0}[ie^2\omega\Pi(q,\omega)/q^2]$, the real part of the polarizability for $v_Fq \ll \omega$ and $\omega \ll \varepsilon_F$ is 
$\mathrm{Re}\Pi(q,\omega) =(2\varepsilon_F\mathcal{\tilde{D}}/\pi)(q/\omega)^2$.  The renormalized plasmon frequency then  
follows by solving $\epsilon(q,\omega) = 1-V_q\mathrm{Re}\Pi(q,\omega) = 0$.  For BLG we find that 
$\omega_{\mathrm{p}}^2 = 4e^2\varepsilon_F\mathcal{\tilde{D}}q$. 
Fig.~\ref{fig3}A shows renormalized Drude weights $\mathcal{\tilde{D}}$ from the full numerical calculations described below. 

\noindent{\em Optical Conductivity}---The full $\sigma(\omega)$ at arbitrary interaction strength is obtained by solving the coupled integral equations Eqs.~(\ref{eq7})-(\ref{eq8}) numerically, 
letting $\omega \to \omega+i\tau^{-1}$.  We can approximately account for screening corrections to our first order interaction self-energy by replacing the bare Coulomb interaction by its 
Thomas-Fermi (TF) statically screened counterpart.   
\begin{figure}[!tb]
  \includegraphics[width=8.5cm,angle=0]{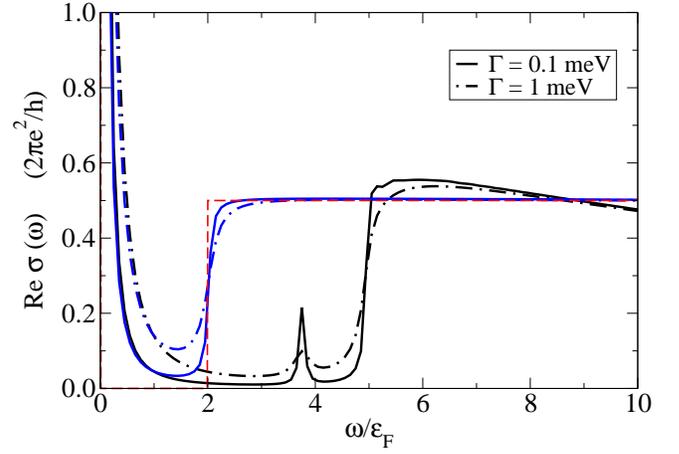} 
\caption{(Color online) Real part of the optical conductivity $\mathrm{Re}\sigma(\omega)$ {\em vs.} frequency $\omega$ at density $n = 10^{12}\mathrm{cm}^{-2}$ for unscreened (black) and screened (blue/grey) e-e interaction. Results for screening with $\kappa = 1$ (corresponding to suspended bilayer graphene) and $\kappa = 4$ (corresponding to a bilayer on a SiO$_2$ substrate) are quantitatively similar and the blue lines show the results for $\kappa = 1$. The disorder broadening is taken as $\Gamma \equiv 1/\tau = 0.1\,\mathrm{meV}$ (solid line) and $1\,\mathrm{meV}$ (dot-dashed line).
The ideal non-interacting case is plotted as the dashed thin red line.} \label{fig2}
\end{figure}
When screening is neglected, interactions 
significantly alter $\sigma(\omega)$ in three respects, as illustrated in Fig.~\ref{fig2}.
First, the interband absorption threshold is changed dramatically from $\omega = 2\varepsilon_F$ to $\omega \simeq 5\varepsilon_F$.  
This effect is analogous to band gap renormalization in regular semiconductors, 
with the Fermi level playing the role of a gap because of Pauli blocking.  The threshold shift is equal to $\Sigma_{k_F+}-\Sigma_{k_F-}$ (Fig.~\ref{fig3}B). 
Second, the value of $\sigma(\omega)$ is no longer universal above the absorption threshold; instead it shows a decreasing trend with $\omega$, first reaching above and then dropping below the non-interacting value $e^2/2\hbar$. Third, an absorption peak appears below the threshold. 
This {\em Mahan exciton} \cite{MahanEx1} feature is a well-understood artifact of our simple self-energy approximation.  
Because of electron scattering processes (including Fermi surface fluctuations due to intraband electron-hole excitations \cite{Nozieres1} and impurity scattering) 
in the conduction band not captured by our self-energy approximation, the Mahan exciton is invariably unstable.
In Fig.~\ref{fig2} we illustrate broadening of the Mahan exciton due to disorder. 

The results obtained when we screen the interactions in our self-energy expression using a TF approximation are shown in grey in Fig.~\ref{fig2}. 
The TF screening wavevector for BLG is given by $q_{\mathrm{TF}} = 4me^2/\kappa$, 
a constant independent of electron density. Surprisingly, we find that with screening (1) the interband absorption threshold shift
nearly disappears, (2) $\mathrm{Re}\sigma(\omega) \simeq e^2/2\hbar$ above the threshold, and (3) the Mahan exciton bound state vanishes. 
In short, the optical conductivity $\mathrm{Re}\sigma(\omega)$ behaves essentially like that of a non-interacting system.

To shed light on this result, we observe that the TF wavevector $q_{\mathrm{TF}} \simeq 2.62 \times 10^9\mathrm{m}^{-1}/\kappa$ for BLG is extremely large compared to all momentum scales of electronic transitions and is, in fact, greater than the momentum cut-off $k_{\mathrm{c}} = \sqrt{2m\gamma_1}$ for both suspended ($\kappa = 1$) and substrate-mounted ($\kappa = 4$) bilayers. 
In a TF screening approximation: $V_q \simeq 2\pi e^2/q_{\mathrm{TF}} \equiv V_0$ in the regime of interest $\varepsilon_F,\omega < \gamma_1$, with the consequence that in Eqs.~(\ref{eq9})-(\ref{eq12}) the band renormalization $\Sigma_{k+}-\Sigma_{k-} \propto V_0\int\mathrm{d}\phi_{k'k}\mathrm{cos}\,2\phi_{k'k}$, and the non-equilibrium self-energies $\delta \Sigma^{z} \propto V_0\int\mathrm{d}\phi_{k'k} \mathrm{cos}\,\phi_{k'k}$, $\delta \Sigma^{\phi,1} \propto V_0\int\mathrm{d}\phi_{k'k}\mathrm{cos}\,\phi_{k'k}\,\mathrm{cos}\,2\phi_{k'k}$, and 
$\delta \Sigma^{\phi,2} \propto  V_0\int\mathrm{d}\phi_{k'k}\mathrm{sin}\,\phi_{k'k}\,\mathrm{sin}\,2\phi_{k'k}$ all vanish.  Strong screening in BLG restores the optical conductivity essentially to its non-interacting value. This remarkable result is peculiar to BLG, since its double-chirality gives rise to spinors with s-wave and d-wave components (rather than s-wave and p-wave as in SLG), which do not couple to p-wave optical dipole transitions through an s-wave short-range interaction. 

Finally we comment on the experimental implications of our findings.  We interpret the weak experimental \cite{WangBLG,BasovBLG,KuzmenkoBLG,CrommieBLG} 
absorption threshold features as evidence  for short-range screened e-e interactions.  
We recognize that the static screening we use could overstate the reduction 
in interaction range and that interaction effects are likely to persist to some degree, 
especially in suspended bilayers for which the dielectric-environment portion of the screening is absent.  
Interaction effect could be 
identified experimentally via the $\omega_{\mathrm{p}}$ renormalizations we predict, for example using electron 
energy loss spectroscopy studies of suspended samples.  Interaction effects might also be more pronounced in 
transitions between the bands near the Fermi energy -- which are included in the massive chiral fermion model we employ -- and the remote bands located approximately $\gamma_1$ away from the Fermi energy. Our theoretical results follow principally from pseudospin chirality, and
will not be strongly influenced by an external potential which opens up a gap at the
Fermi level of an undoped system.  We therefore can expect on the basis of current
results, that excitonic binding energies in these gapped systems will be suppressed. 

In conclusion, we have developed a theory for the e-e interaction effects on the optical conductivity $\sigma(\omega)$ of doped bilayer graphene. We find a novel coupling effect 
which couples the Drude and interband response of the optical conductivity, and an accompanying renormalization of the leading-order plasmon frequency. We also find that screening dramatically suppresses band renormalization and excitonic effects, restoring $\sigma(\omega)$ very close to the universal value $e^2/2\hbar$ above the absorption threshold.  

W.-K. thanks Qian Niu for useful discussions. This work was supported by the Welch Foundation and by the DOE.


\begin{thebibliography}{20}
\bibitem{gapless} A gap can be induced in bilayer graphene by an external electric field which produces an electric potential difference 
between the two layers and breaks inversion symmetry.  See for example Refs.~[\onlinecite{CrommieBLG,KusmenkoNewPreprint}] and 
work cited therein.  In this article we focus on the balanced bilayer case in which inversion symmetry is maintained.
\bibitem{interactionapology}  We are neglecting the small difference in interaction energy between electrons in the 
same layer and electrons in different layers.
\end{thebibliography}
\end{document}